\documentclass[a4paper]{jpconf}
\pdfoutput=1
\usepackage[utf8]{inputenc}
\usepackage[english]{babel}
\usepackage{multirow}
\usepackage{graphicx}
\usepackage[numbers,sort&compress]{natbib}
\usepackage[usenames,dvipsnames,x11names]{xcolor}

\definecolor{dblue}{rgb}{0,0,0.4} 
\definecolor{dblue4}{rgb}{0.06,0.31,0.55} 
\usepackage[colorlinks=true,urlcolor=dblue4,anchorcolor=dblue4,citecolor=dblue4,filecolor=dblue4
,linkcolor=dblue4,menucolor=dblue4,pagecolor=dblue4,linktocpage=dblue4,pdfa=true]{hyperref}

\usepackage[activate={true,nocompatibility},final,tracking=true,kerning=true,spacing=true,factor=1100,stretch=10,shrink=10]{microtype}

\usepackage{txfonts}
\usepackage[T1]{fontenc}

\begin{document}
\title{\textcolor{dblue}{Radiative neutrino mass generation from WIMP dark~matter}}

\author{\href{http://goo.gl/00TnL}{Roberto A. Lineros}}

\address{Instituto de F\'{\i}sica Corpuscular -- CSIC/U. Valencia,\\ Parc Cient\'{\i}fic, calle Catedr\'{a}tico Jos\'{e} Beltr\'{a}n 2, E-46980 Paterna, Spain}

\ead{rlineros@ific.uv.es}

\begin{abstract}
The minimal seesaw extension of the Standard Model requires two electroweak singlet fermions in order to accommodate the neutrino oscillation parameters at tree level. Here we consider a next to minimal extension where light neutrino masses are generated radiatively by two electroweak fermions: one singlet and one triplet under SU$(2)$. These should be odd under a parity symmetry and their mixing gives rise to a stable weakly interactive massive particle dark matter candidate. For mass in the GeV-TeV range, it reproduces the correct relic density, and provides an observable signal in nuclear recoil direct detection experiments. The fermion triplet component of the dark matter has gauge interactions, making it potentially detectable at present and near future collider experiments.
\end{abstract}

\section{Introduction}
\label{sec:intro}
It is well established that most of the matter content of the Universe is in the form of Dark Matter~(DM).
The latest observations of the Cosmic Microwave Background anisotropies~\cite{2015arXiv150201589P} implies that the DM relic abundance in the $\Lambda$CDM model is
\begin{equation}
\label{eq:ra}
\Omega_{\rm DM}h^2 = 0.1198 \pm 0.0015 \, ,
\end{equation}
where $h = 0.678 \pm 0.009$ is the scale factor for Hubble expansion rate~\cite{2015arXiv150201589P}.
The value of the relic abundance can be explained in terms of Weakly Interactive Massive Particles~(WIMPs).
This class of particles are not part of the Standard Model~(SM) indicating the necessity to go beyond.
The existence of DM motivates, in part, searches of new particles in particle accelerator like the Large Hadron Collider~(LHC).

In general words, WIMP-DM particles can easily reproduce the observed value of the relic abundance thanks to the \emph{freeze-out mechanism}.
This mechanism comes from the combined effect of the expansion of the Universe and the interaction rate among DM and the primordial plasma during the early Universe.

For WIMP particles in GeV--TeV range, the relic abundance is 
\begin{equation}
\Omega_{\rm WIMP}h^2 \simeq 0.1 \frac{3 \times 10^{-26} \, {\rm cm}^3/{\rm s}}{\langle \sigma v \rangle} \, ,
\end{equation}
where $\langle \sigma v \rangle$ is the DM thermally averaged cross section.
The correct value of $\Omega_{\rm DM}h^2$ is then obtained when $\langle \sigma v \rangle$ is of the order of weak interaction cross sections i.e. $3 \times 10^{-26} \, {\rm cm}^3/{\rm s}$.
In minimal scenarios, DM models are constructed on top of the SM providing also new mediators to act as portal between the dark and visible sectors.
However, it seems reasonable that DM is part of dark sector that includes new particles and interactions like in Supersymmetric, extra dimensions~\cite{2014JCAP...10..059L}, or dark gauge group models~\cite{2013PhRvD..87l3521F}.\\

On the other hand, neutrinos also give an indication that the SM is not complete.
In the SM, neutrinos are massless but the observation of neutrino oscillations indicate the contrary, i.e. they are massive and with the largest mass squared differences of $\sim 10^{-3}\,{\rm eV}^2$~\cite{Schwetz:2011zk,Tortola:2012te}.
The most known mechanism of neutrino mass generation are the so called \emph{see-saw} models (see for instance~\cite{2014NJPh...16l5012L} and references within), where the smallness of neutrino masses is generated at tree level by extremely heavy states such in the case of the Type-I see-saw.
However, it is also possible to generate the neutrino masses via loop processes.
One realization are the \emph{scotogenic} models~\cite{Ma:2006km,Ma:2008cu}.
In one of its versions, the SM is enlarged with an extra SU$(2)_L$ doublet scalar and a majorana fermion singlet.
In another version, it gets an extra SU$(2)_L$ doublet scalar but a SU$(2)_L$ fermion triplet with zero hypercharge.
In this type of models, an ad-hoc symmetry is imposed and therefore neutrino mass generation at tree level is forbidden.
This scheme is very tantalizing due to that the ad-hod symmetry allows to have a DM candidate.

Our model~\cite{2013JHEP...10..149H} is based on a mixed realization of the scotogenic models.
We show that in this scheme the phenomenology is richer and allows to have a WIMP-DM candidate with a sizeable interaction with quarks and a mass ranging from GeV to TeV.\\

\section{The model}
\label{sec:model}

As previously stated, the model is a mix realization of two scotogenic models.
Besides the particle content of the SM, we have included a majorana fermion singlet $N$, a SU$(2)_{L}$ fermion triplet $\Sigma$, a SU$(2)_{L}$ scalar triplet $\Omega$; the latter three are without hypercharge, and finally a SU$(2)_{L}$ scalar doublet $\eta$.
The new fields, except $\Omega$, are charged with a $Z_2$ symmetry that ensures the DM stability.
Due to $\Omega$ is not protected by the symmetry, it will participate in the spontaneous symmetry breaking of the SU$(2)_L \times$~U$(1)_Y$ SM group.
All the matter content of the model is summarized in Tab.~\ref{tab}.
\begin{table}[t]
\centering
\begin{tabular}{|c||c|c|c||c|c||c|c|}
\hline
        & \multicolumn{3}{|c||}{Standard Model} &  \multicolumn{2}{|c||}{Fermions}  & \multicolumn{2}{|c|}{Scalars}  \\
        \cline{2-8}
        &  $L$  &  $e$  & $\phi$  & $\Sigma$ &  N   & $\eta$ & $\Omega$ \\
\hline                                                                  
SU$(2)_L$ &  2    &  1    &    2    &     3    &  1   &    2   &    3     \\
U$(1)_Y$     & -1    &  -2    &    1    &     0    &  0   &    1   &    0     \\
$Z_2$   &  $+$  &  $+$  &   $+$   &    $-$   & $-$  &   $-$  &   $+$    \\
\hline
\end{tabular}
\caption{Matter content.}
\label{tab}
\end{table}

Using the symmetries described in the matter content, we write the lagrangian of the model
\begin{eqnarray}
\mathcal{L} &\supset& - Y_{\alpha \beta}\,\overline{L}_{\alpha}  e_{\beta} \phi- Y_{\Sigma_\alpha} \overline{L}_{\alpha}C\Sigma^{\dagger} \tilde{\eta} - 
\frac{ 1 }{4}M_\Sigma \mbox{Tr} \left[\overline{\Sigma}^{c} \Sigma \right] + \nonumber \\
& & -Y_{\Omega} \mbox{Tr}\left[ \overline{\Sigma} \Omega \right]N - Y_{N_\alpha} \overline{L}_{\alpha}\tilde{\eta} N - \frac{1}{2}M_N \overline{N}^{c}N 
+ h.c.\, - V_{\rm scal}, 
\label{eq:lagrangian}
\end{eqnarray}
where $Y_X$ stand for yukawa couplings, $M_X$ are mass terms of $\Sigma$ and $N$, and $V_{\rm scal}$ is the scalar potential.
The details about the scalar potential are in Ref.~\cite{2013JHEP...10..149H}.\\

The neutrino masses arises at one loop mainly due to the terms $L \eta N$ and $L \Sigma \eta$ present in the lagrangian.
We obtain that the neutrino mass matrix in the flavor basis is
\begin{equation}
M_{\alpha \beta}^{\nu} = \sum_{\sigma=1,2} \frac{h_{\alpha\sigma} h_{\beta\sigma}}{8 \pi^2} \frac{\lambda_5 v_h^2}{m_0^2} M_k \, ,
\end{equation}
where $h_{\alpha \sigma}$ is a $3\times2$ matrix that encodes the values of the yukawa terms $Y_{\Sigma}$ and $Y_{N}$. $\lambda_5$ and $m_0$ are terms coming from the scalar sector. $v_h$ is the vacuum expectation value of the doublet $\phi$, and $M_k$ the masses of the mass eigenstates from the mix of $N$ and $\Sigma^0$ (neutral component of the triplet $\Sigma$).\\

The model contains two possible DM candidates.
One is the neutral component of the scalar doublet $\eta$ which has a similar phenomenology to the inert higgs model; and the second is the fermion which is a combination among $N$ and $\Sigma^0$.
In our analysis, we focus on the fermion DM candidate.
The two neutral fermions mass eigenstate are combination of $N$ and $\Sigma^0$ and we will denote as $\chi^0_{1,2}$.
This class of mixed state, between singlet and triplet, was not possible in the traditional versions of the scotogenic model.
In our model, this is possible thanks to $\Omega$ that allows the term $\Omega \Sigma N$ in the lagrangian. 
This term also generates a mass splitting between the neutral and charged component of the triplet $\Sigma$ which later it will reveal to be essential for the DM phenomenology.\\

\begin{figure}[t]
\centering
\includegraphics[width=0.7\textwidth]{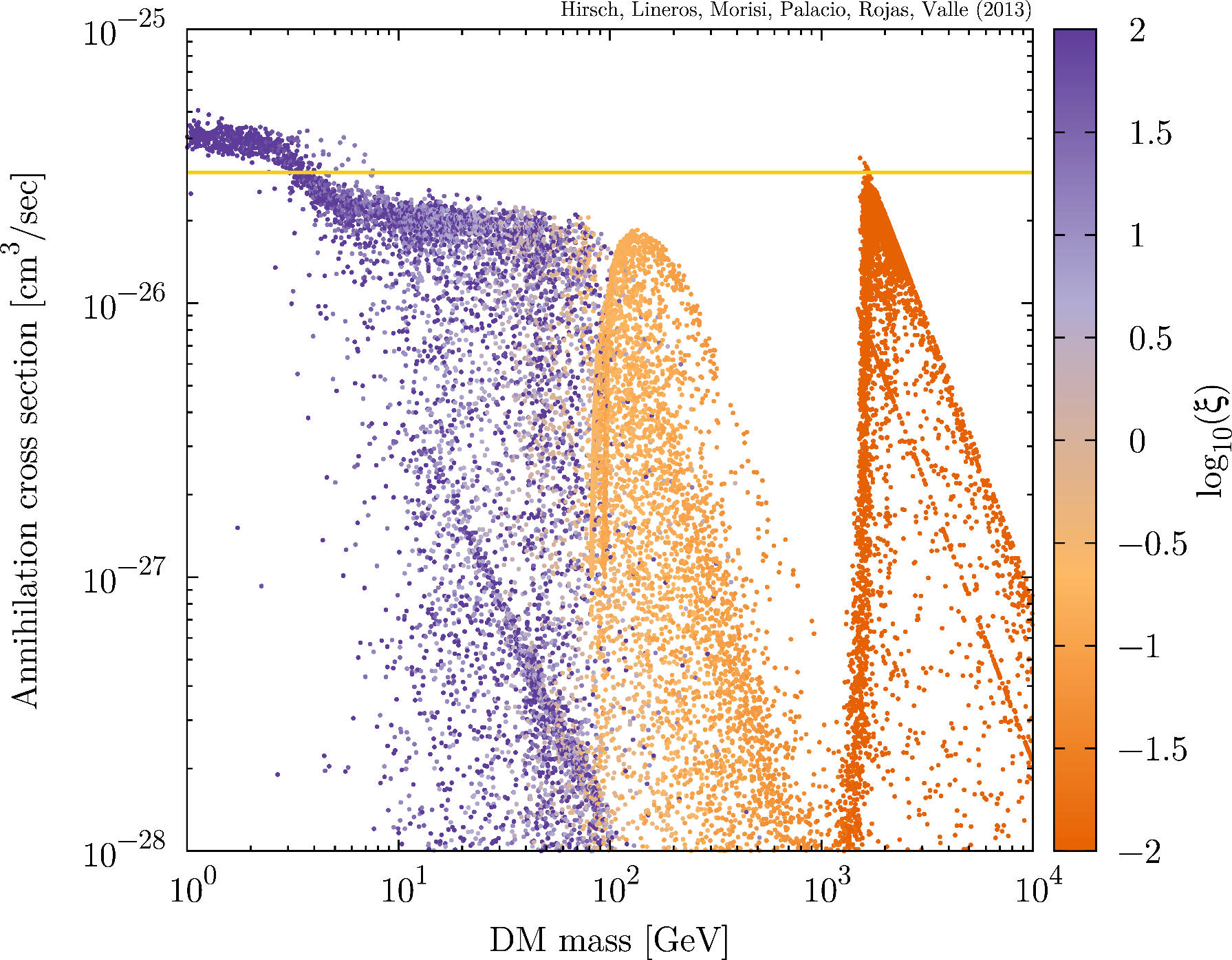}
\caption{Annihilation cross section versus DM mass. More details in Ref.~\cite{2013JHEP...10..149H}}
\label{fig1}
\end{figure}
At this point, it is important to describe the mass spectrum of the fermion states:
\begin{eqnarray}
m_{\chi^{\pm}} &=& M_{\Sigma} \, , \\
m_{\chi^{0}_1} &=& \frac{1}{2} \left( M_{\Sigma} + M_{N} - \sqrt{ \left(M_{\Sigma}-M_{N}\right)^2 + 4 \left( 2 Y_{\Omega} v_{\Omega}\right)^2}\right) \, ,\\
m_{\chi^{0}_2} &=& \frac{1}{2} \left( M_{\Sigma} + M_{N} + \sqrt{ \left(M_{\Sigma}-M_{N}\right)^2 + 4 \left( 2 Y_{\Omega} v_{\Omega}\right)^2}\right) \, , \\
\tan(2\alpha) &=& \frac{4 Y_{\Omega} v_{\Omega}}{M_{\Sigma} - M_{N}} \, ,
\end{eqnarray}
where $v_{\Omega}$ is the vacuum expectation value of the neutral component of $\Omega$ and $\alpha$ is the mixing angle between $N$ and $\Sigma^0$.
Here it is easy to appreciate the role of $v_{\Omega}$ breaking the degeneracy between $\chi^{\pm}$ and the $\chi^0_i$ with larger component of $\Sigma^0$.\\

We use the codes {\tt LanHEP}~\cite{lanhep:1996,lanhep:2009,lanhep:2010} and {\tt MicrOmegas}~\cite{micromegas:2013} to scan on the parameter space (See details in Ref.~\cite{2013JHEP...10..149H}) including constraints from neutrino masses, perturvativity limits of the adimensional parameters, and LEP~\cite{L3:2001PhLB} and CMS~\cite{CMS:2012PhLB} searches on new charged particles.
Also, we require that $m_{\chi^0_1} < m_{\chi^\pm}, m_{\eta}$.
We obtained the parameter space compatible with current relic abundance value (Eq.~\ref{eq:ra}).
The allowed region is compatible with DM masses in the range of $10$ to $10^4$~GeV.\\

In Fig.~\ref{fig1}, we present the result of the scan in terms of the annihilation cross section (relevant for indirect detection searches) versus the DM mass.
The color code corresponds to the value of 
\begin{equation}
	\xi = \frac{M_{\Sigma} - m_{\rm DM}}{m_{\rm DM}} \, ,
\end{equation} 
which is related to the amount of $\Sigma^0$ present in DM candidate.
This quantity reveals three regimes.
The low DM mass corresponds to a DM which is mainly singlet $N$.
This is the consequence of the LEP constraints on charge particles.
At the TeV masses, the DM candidate is mostly triplet.
This is produced due to annihilation channels to gauge bosons, coannihilation with $\chi^{\pm}$, and $\chi^{\pm}$ annihilations.
In the range between 100~GeV to 1~TeV, the DM is more a mixed state $N$-$\Sigma^0$.
This regime is very interesting because annihilations into quarks become relevant.
This occurs because $\Omega \Sigma N$ allows a higgs portal.\\

\begin{figure}[t]
\centering
\includegraphics[width=0.7\textwidth]{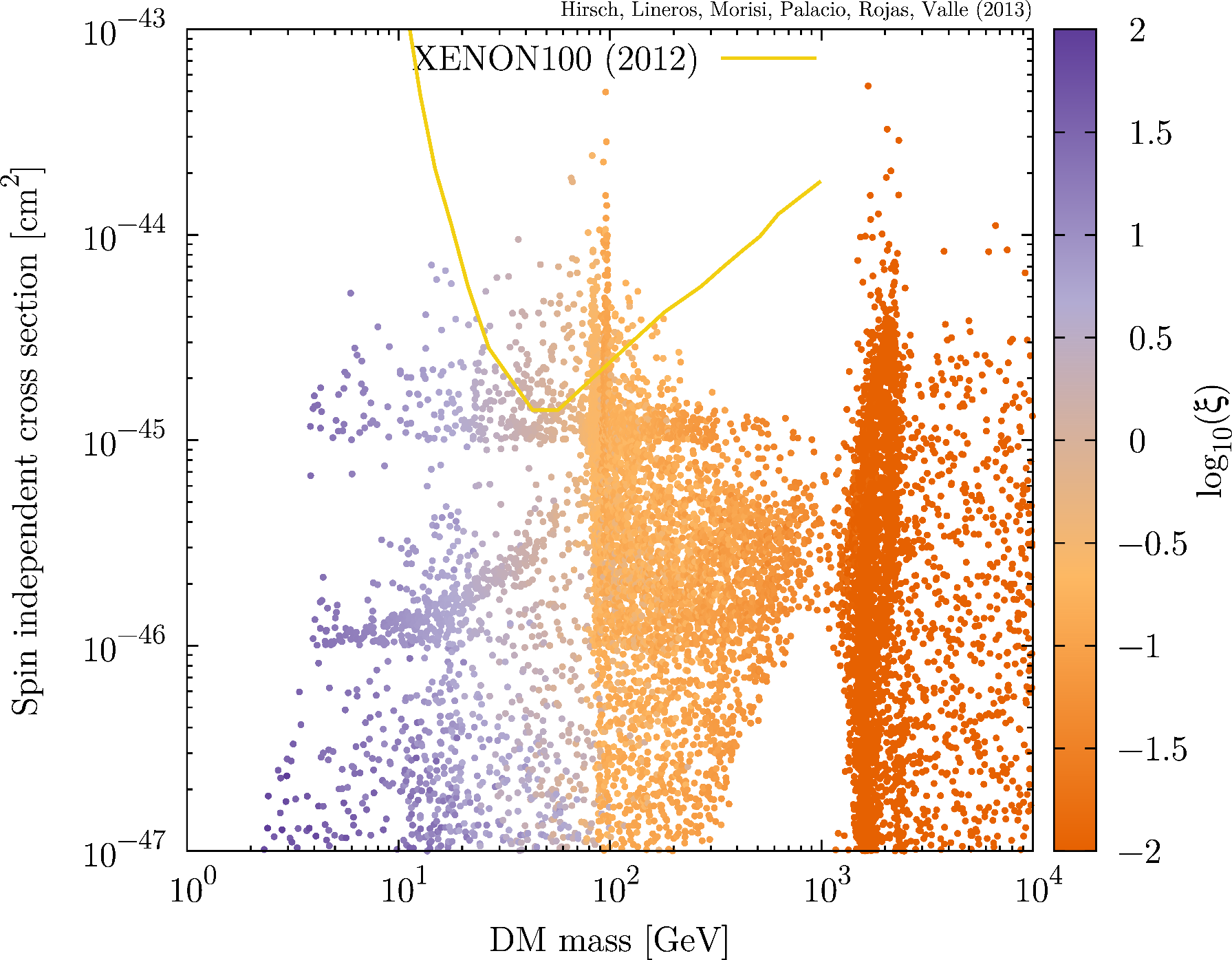}
\caption{Spin-independent cross section versus DM mass. More details in Ref.~\cite{2013JHEP...10..149H}}
\label{fig2}
\end{figure}

Same three regimes can be observed in Fig.~\ref{fig2} where we show the spin-independent cross section versus DM mass.
Let us highlight that in traditional scotogenic models the expected spin-independent cross section is largely suppressed because DM can only interact with quarks via loops.
However, in our model, the direct detection signal is of the order of the reach of current experiments thanks to the higgs portal produced by the term $\Omega \Sigma N$.
In Fig.~\ref{fig2}, we also show latest exclusion limits of XENON100~\cite{XENON:2012Ph} as reference.\\

\section{Conclusions}
\label{sec:concl}

The nature behind DM is still a mystery.
WIMPs are the most attractive candidates due to many search strategies. 
We have proposed a new scotogenic model that provide a neutrino mass mechanism at one loop and a fermion WIMP-DM candidate.
The fermion DM in this model provides features that are not present in the traditional scotogenic models.
One feature is the "tree level" spin independent cross section that is at the reach of current and future direct detection experiment.
Also, the DM mass can go from 10~GeV to 10~TeV which is broader with respect to previous models.\\

\section*{Acknowledgments}
We would like to thank the organizers of the TAUP 2015 for an interesting conference.
This work was supported by the Spanish MINECO under grants FPA2014-58183-P, and MULTIDARK CSD2009-00064 (Consolider-Ingenio 2010 Programme); by Generalitat Valenciana grant PROMETEOII/2014/084, and Centro de Excelencia Severo Ochoa SEV-2014-0398.
\href{http://goo.gl/00TnL}{R.~L.} is supported by a Juan de la Cierva contract JCI-2012-12901 (MINECO).

\def\apj{Astrophys.~J.}                       
\def\apjl{Astrophys.~J.~Lett.}                
\def\apjs{Astrophys.~J.~Suppl.~Ser.}          
\def\aap{Astron.~\&~Astrophys.}               
\def\aj{Astron.~J.}                           %
\def\araa{Ann.~Rev.~Astron.~Astrophys.}       %
\def\mnras{Mon.~Not.~R.~Astron.~Soc.}         %
\def\physrep{Phys.~Rept.}                     %
\def\jcap{J.~Cosmology~Astropart.~Phys.}      
\def\jhep{J.~High~Ener.~Phys.}                
\def\prl{Phys.~Rev.~Lett.}                    
\def\prd{Phys.~Rev.~D}                        
\def\nphysa{Nucl.~Phys.~A}                    

\bibliographystyle{iopart-num}
\bibliography{lineros-taup.bib}

\end{document}